# Auditory-visual scenes for hearing research

Steven van de Par[1*], Stephan D. Ewert[1*], Lubos Hladek[2], Christoph Kirsch[1], Julia Schütze[1], Josep Llorca-Bofí[3], Giso Grimm[1], Maartje M.E. Hendrikse[1,4], Birger Kollmeier[1], Bernhard U. Seeber[2]

## Abstract

While experimentation with synthetic stimuli in abstracted listening situations has a long standing and successful history in hearing research, an increased interest exists on closing the remaining gap towards real-life listening by replicating situations with high ecological validity in the lab. This is important for understanding the underlying auditory mechanisms and their relevance in real-life situations as well as for developing and evaluating increasingly sophisticated algorithms for hearing assistance. A range of 'classical' stimuli and paradigms have evolved to de-facto standards in psychoacoustics, which are simplistic and can be easily reproduced across laboratories. While they ideally allow for across laboratory comparisons and reproducible research, they, however, lack the acoustic stimulus complexity and the availability of visual information as observed in everyday life communication and listening situations.

This contribution aims to provide and establish an extendable set of complex auditory-visual scenes for hearing research that allow for ecologically valid testing in realistic scenes while also supporting reproducibility and comparability of scientific results. Three virtual environments are provided (underground station, pub, living room), consisting of a detailed visual model, an acoustic geometry model with acoustic surface properties as well as a set of acoustic measurements in the respective real-world environments. The current data set enables i) audio-visual research in a reproducible set of environments, ii) comparison of room acoustic simulation methods with "ground truth" acoustic measurements, iii) a condensation point for future extensions and contributions for developments towards standardized test cases for ecologically valid hearing research in complex scenes.

Footnote & affiliations:

*)  These authors share first authorship

1)  Carl-von-Ossietzky Universität, Oldenburg, Dept. Medical Physics and Acoustics, Cluster of Excellence "Hearing4all"
2)  Technical University of Munich, Audio Information Processing, Department of Electrical and Computer Engineering
3)  RWTH Aachen University, Institute for Hearing Technology and Acoustics
4)  Erasmus University Medical Center, Rotterdam, Department of Otorhinolaryngology and Head and Neck Surgery

# 1. Introduction

Speech is an important mode of communication in a wide range of daily-life situations. In real life, however, speech communication is often challenged by a number of complicating factors, such as the presence of ambient, interfering noise, a dynamically changing acoustic environment, the presence of reverberation, but also due to individual difficulties in understanding speech caused by hearing impairment (e.g., Plomp, 1978) or due to unfamiliarity with the specific language (e.g., Warzybok et al., 2015).

In order to get a better understanding of the challenges that these complicating factors impose on persons with a reduced speech recognition ability (e.g., hearing impaired listeners), much prior research has focussed on conditions with relatively simple, artificial stimuli. For example, balanced speech corpora (e.g., rhyme tests or matrix sentence tests) have been employed in conditions with one or two interfering sources like speech shaped noise played over a few loudspeakers or headphones. A particular advantage of using these artificial stimulus conditions is that stimulus properties are well defined and dedicated strategies can be employed to investigate how a certain stimulus property affects speech intelligibility. This helps to understand the mechanisms underlying speech processing in humans, and allows developing models of speech intelligibility (e.g. ANSI 1997, Steeneken and Houtgast, 1980; Rhebergen et al., 2006; Beutelmann et al., 2009; Jørgensen et al., 2013; Biberger and Ewert, 2016). Moreover, stimuli and experimental methods as well as model predictions can be reproduced and compared across different labs, which generally contributes to scientific development.

However, the ecological validity of the described classical "lab-based" speech and psychoacoustical experiments using such artificial stimulus conditions has been questioned with regard to real-world outcomes (e.g., Kerber & Seeber, 2011; Cord et al., 2007; Jerger, 2009; Miles et al., 2020). For noise reduction schemes and beamforming algorithms in hearing aids it has indeed been shown that there is a discrepancy between laboratory results obtained with simple speech intelligibility measurements and real-life results (Bentler, 2005; Cord et al., 2004). To close the gap towards real life listening, tests need to consider additional factors that may affect speech perception in everyday listening situations (for a review see Keidser et al., 2020). Keidser et al. provided a comprehensive set of such factors which were grouped in five methodological dimensions: 'Sources of stimuli, Environment, Context of participation, Task, and Individual'.

Beyond speech intelligibility, virtual acoustic and audio-visual environments can help us to better understand binaural hearing in reverberant spaces and the ability to locate sound sources (e.g. Seeber and Clapp, 2020), the mechanisms of auditory scene analysis, the quality of sound reproduction achieved with ear-level devices or in a to-be-built concert hall, or the mechanisms of shared attention between acoustic and visual sources. Like for speech research, the complexity of the acoustic scene and the audio-visual configuration will affect performance.

Real-life acoustical *Environments* are typically more complex than classical "lab-based" speech intelligibility tests using speech in (stationary) noise. Such complex acoustic environments (e.g., Weisser et al., 2019) typically contain multiple, diverse spatially distributed interfering *Sources of stimuli* such as speech, music, and diffuse background noise. Moreover, reverberation is typical in enclosed spaces. Numerous studies have assessed the effect of specific aspects of interferers and their spatial distribution on speech intelligibility. It is known that the spectro-temporal properties of interfering sources influence speech intelligibility. For example, when a few interfering speakers (fluctuating interferers) are employed, listening into the temporal gaps improves intelligibility (e.g.,

Brungart and Iver, 2012; Lingner et al., 2016). The spatial separation between interfering speech and attended speech improves intelligibility (e.g, Bronkhorst and Plomp, 1988; Peissig and Kollmeier, 1997; Best et al., 2015; Schoenmaker et al., 2016; Ewert et al., 2017). This spatial benefit in speech intelligibility, however, is reduced in reverberant environments and may strongly depend on the orientation and position with respect to reflecting surfaces nearby (Beutelmann and Brand, 2006; Biberger and Ewert, 2019). Depending on the specific real-life acoustic environment, these effects can be assumed to occur in specific combinations and influence speech intelligibility in a particular manner. Therefore, the development of complex acoustic environments representative of a large variety of real-life everyday environments can be highly relevant for obtaining ecologically valid estimates of speech intelligibility. In addition, the assessment of the effectiveness of algorithms for hearing devices in real-life may depend on having a realistic audio-visual environment that provides a well-defined *Context of participation* which can elicit natural behaviour of the participant such as for example head movements. Several studies have shown that "knowing where to attend to", i.e., a predictable as opposed to an unpredictable stimulus location, can improve speech intelligibility (Kidd, et al., 2005; Teraoka et al., 2017; Schoenmaker et al., 2016). Visual cues can guide the spatial attention of the listener (Best et al., 2007; Hendrikse et al. 2020) and can affect the self-motion of listeners, which can in turn influence speech intelligibility, e.g., as a consequence of altered head orientation (Grange & Culling, 2016). Moreover, lip reading, or speech reading, can specifically contribute to speech intelligibility in noisy situations (Sumby and Polack, 1954; MacLeod and Summerfield, 1987; Schwartz et al., 2004).

One option to adopt these aspects naturally lies in field tests in real daily-life situations. However, in contrast to laboratory experiments, control over the precise acoustic condition and stimulus properties is typically very limited which might in turn affect evaluation of the results and the development of auditory models. An alternative are spatial and dummy head recordings of realistic scenes (e.g., Kayser et al., 2009; Grange and Culling, 2016; Culling, 2016), allowing for the reproduction of existing acoustic scenes in the laboratory, however, with limited flexibility regarding the controlled modification of the scenes as well as interactive behaviour in the scene, such as natural head movements.

One more recent option is the use of virtual (acoustic) environments (VE) to produce realistic audio-visual scenarios for which all properties can be measured and controlled (e.g., Kerber & Seeber, 2013; Pausch et al., 2018; Blau et al., 2021). Virtual reality techniques which synthesize the acoustic scene by (room) acoustics simulation and render the visual scene using computer graphics allow to systematically manipulate and interact with the scene. Several systems for (room) acoustics simulation and auralization with and without a visual component exist (e.g., Hafter & Seeber, 2004; Seeber et al., 2010; Schröder and Vorländer, 2011; Wendt et al., 2014; Grimm et al., 2016). If a high degree of realism is reached, virtual audio-visual environments thus offer the opportunity to precisely control and reproduce certain stimulus properties to, e.g., probe hypotheses about auditory processing, and to test the effectiveness of hearing aid algorithms, while at the same time reaching a high degree of ecological validity, ideally exactly as in the corresponding real-life scenario.

For classical psychoacoustic experiments, certain methodologies have been widely used across labs (e.g. transformed up-down, Levitt 1971; matrix sentence test, Hagermann, 1982; Wagener et al., 1999; Kollmeier et al., 2015), which, combined with the acoustic calibration of broadly similar loudspeaker or headphone setups, leads to comparable results. Such established methods and measures, forming "de-facto" standards and enabling comparison of results across different research sites, do not yet exist for virtual audio-visual environments with applications in hearing research. Especially, considering the higher complexity of VR systems and particularly the differences in setups and acoustic and visual rendering techniques, the degrees of freedom in which different lab solutions for the same problem might deviate are considerably increased. As a consequence, reproducibility and comparability across labs requires a special effort.

For this reason, this contribution presents a framework for defining and documenting complex audio-visual environments and embedded realistic communication "scenes" for hearing research, with the aim to stipulate increased reproducibility and comparability of research across labs. In this context, the environment refers to a specific audio-visual surrounding typically encountered in real life, like a living room. The term "scene" refers to a specific (communication) situation in a given environment, e.g., a conversation between two people seated on sofa chairs in the living room. Within an environment, a multitude of scenes can be defined. The proposed framework defines the required formats in which information about the environment and the scene needs to be provided to i) enable recreation in different labs and to ii) be extendable with further environments and scenes for new experiments. For visual and acoustic rendering of the environment, a geometric model provides detailed information for the visual representation, and coarser information for the acoustic representation. Simple (albedo) textures define visible surfaces, while acoustic surfaces are characterized by their absorptive and scattering properties. These definitions should be independent of the systems they are rendered with in order to provide greatest flexibility for use in different laboratories and over time.

Within the proposed framework, three example audio-visual environments (an underground station, a pub, and a living room) are specified according to the suggested format and supplemented with acoustic "ground truth" measurements obtained in the corresponding real-life environments. These three environments represent relevant daily-life situations in which understanding speech may be challenging.

Within these three environments, two scenes each define representative source-receiver positions and orientations. One scene is motivated by audiological standard-test distribution of sources as far as reasonable in the context of the environment, while the other scene varies additional parameters, e.g., distances or interferer positions. Independent variables are the angular position and distance of sources. In addition, representative interfering source positions and/or signals are provided fitting the context of the environment. To verify acoustic and visual rendering methods, three types of acoustic measurements (omni-directional, Ambisonics, and dummy head recordings) of selected source-receiver combinations from the existing real-life spaces used for the scenes are provided. Still images are provided for visual verification.

The current contributions with documentations are hosted in a dedicated channel on the open Zenodo platform ([https://zenodo.org/communities/audiovisual_scenes/](https://zenodo.org/communities/audiovisual_scenes/)), and new contributions from the community are cordially invited. With the provided information, the current contributions should enable researchers to reproduce the same virtual environments using their preferred visual and acoustical rendering methods. Based on future evaluation of the suggested and additionally contributed environments across different research laboratories, this contribution can serve as a condensation point towards standardized test cases for ecologically valid hearing research in complex scenes.

## 2. Audio-visual environments and scenes

Three audio-visual environments (underground station, pub, and living room; left to right in Fig. 1), modelled after real-life enclosed spaces are provided. They cover a large variety of everyday communication situations with different degrees of complexity and reverberation time. For each environment, a visual model is provided (upper row of Fig. 1) as well as at least one simplified geometry model suited for real-time room acoustics simulation (middle row of Fig. 1). In each of the three environments, two audio-visual scenes are defined, being comprised of specific combinations of acoustic sources and receivers, resembling distances and spatial configurations typical for communication in the respective environments. The lower row shows the floor plan for each of the environments including the looking direction (in red). In the first scene, several source positions are arranged in a circular manner around the listener, adapted to the geometry and natural communication distances in the environment. This scene represents a spatial configuration commonly used in audiology research. In the second scene, for a specific angular position, the focus is on different distances to the source appropriate for the environment. For each scene, acoustic measurements in the real-life spaces were performed, so that the scenes can be acoustically recreated using recorded impulse responses from the source position to the receiver or modelled using room acoustics simulation with the acoustic measurements serving as reference. In addition, we also provide room acoustics parameters of the environment.

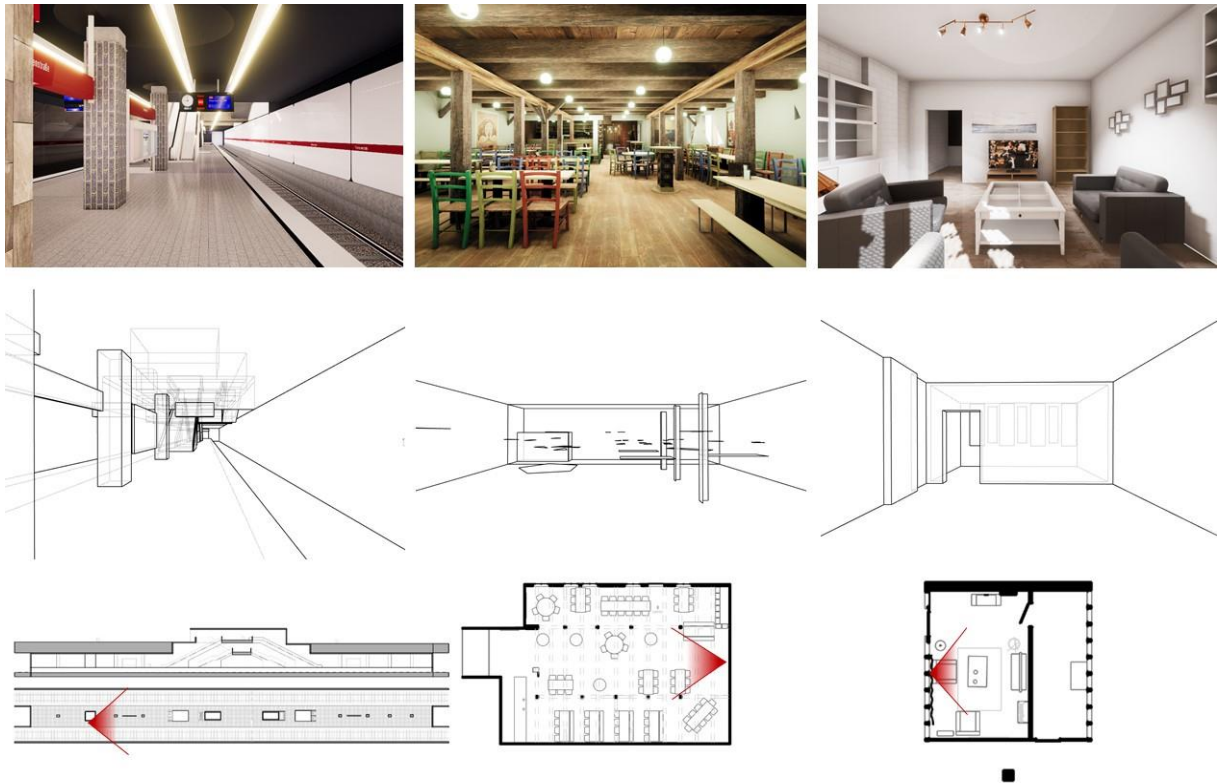

*Figure 1: The three example environments underground station, pub, and living room (left to right) modelled after an underground station in the center of Munich, the OLs Brauhaus in Oldenburg, and the Living Room Laboratory created at the University of Oldenburg for research purposes. Upper row: Visual rendering in Unreal engine. Middle row: Wire frame representation of the simplified geometry for room acoustics simulation. Lower row, the view frustum of the renderings is indicated in the floor plans which are depicted in more detail in Fig. 2-4.*

More detailed descriptions as well as the measurements and models of the environments are provided as part of the freely available dataset structured as described in Sec. 4 (https://zenodo.org/communities/audiovisual_scenes/).

# Underground Station

This environment represents the platform of the underground (U-Bahn) station Theresienstraße in Munich, Germany (see left column of Fig. 1). The detailed floor plan is provided in Fig. 2, showing the strongly elongated and large environment with overall dimensions of 120 m x 15.7 m and a ceiling height of 4.16 m from the platform, extending to 11.54 m around the escalators. The environment involves the lower platform and a part of the upper floor around the escalators. The volume of the platform space is 8555 m3, which increases to 11084 m3 when the area around stairs and escalators is included. The floor, column and track surfaces are composed of hard, acoustically reflective materials (stone tiles, concrete, crushed rock track ballast) while side walls and ceiling are covered with paneling and acoustical treatment. The reverberation time T30 ranges from 2.44 s at 250 Hz to 0.65 s at 8 kHz and the early decay time (EDT) changes between 1.46 s and 0.46 s in the same frequency range.

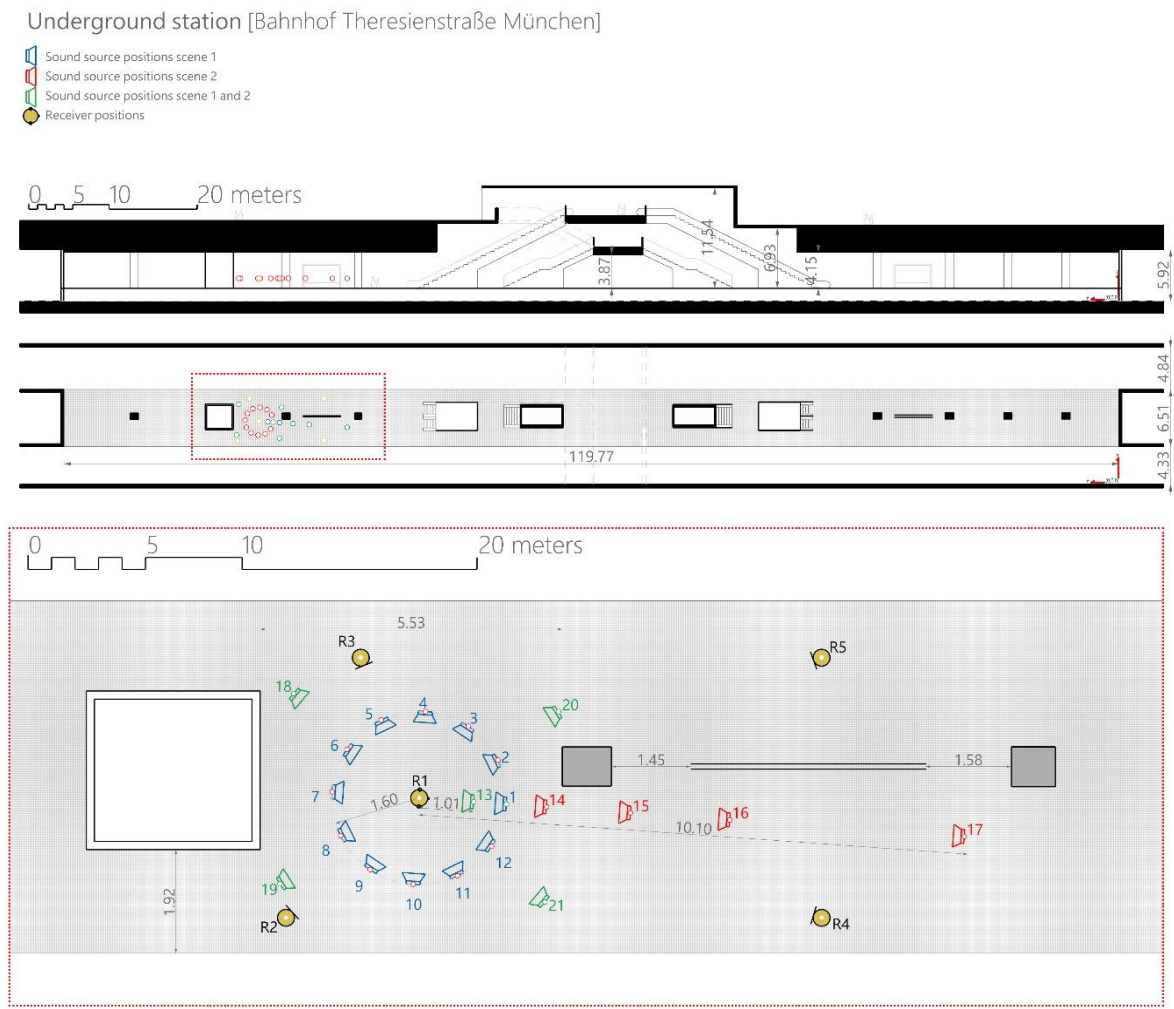

*Figure 2. Upper panels: Cross section (top) and floorplan (middle) of the underground station environment with dimensions in meters. Lower panel: Magnified view of the area with receivers (R1-5, yellow head and microphone symbols) and sources (loudspeaker symbols) indicating their orientation in the horizontal plane. The blue loudspeakers are part of scene 1 and the red loudspeakers of scene 2. Green loudspeakers can be used in both scenes for, e.g., interfering sources.*

The environment is typically noisy, with fans providing air circulation and cooling the video projection system while escalators rumble – in the absence of the noise from incoming trains and announcements. Selected background sounds were recorded individually to recreate the environment's noise with a room acoustic simulation. Additional recordings with a multi-microphone array at the receiver position preserve spatial properties of the sound and can be used to recreate the acoustic background at the receiver position, e.g. with Ambisonics rendering.

Monaural room impulse responses were measured from several source positions to all receiver positions and can be used to verify the acoustic simulation. Multi-channel room impulse responses from various source positions to the listener position R1 can be used to present spatialized sources or interferers to the listener without the use of room simulation techniques.

### Scene 1: Nearby communication – equidistant sources with semi-distal background noise sources

The first scene resembles communication with one or more nearby persons standing on the platform (see lower panel of Fig 2; receiver R1, source positions 1 to 12). The listener is in the center and the talker could stand in one of twelve possible positions equidistantly spaced at 30° around the listener. The positions are in 1.6 m distance from the listener. This arrangement represents frequently used configurations in audiology research. Semi-distal noise sources are distributed at four angles (30°, 150°, 210°, 330°) at a distance of 2.53 m from the receiver, which could be used to create interference from other people or other noise sources at the platform.

### Scene 2: Approaching person – radially spaced sources

The second scene represents a situation where a person is approaching or receding from the listener (see lower panel of Fig 2, receiver R1, source positions 13, 1, 14, 15, 16). The sources are radially distributed along one line at distances from 1 m to 10 m such that the change of the direct sound is 4 dB. Scene 1 and scene 2 are arranged around the same listener position R1 and share source position 1 on the circle in front of the listener and the interferer positions 18-21.

A more detailed description of the underground scene acoustics along with speech intelligibility data obtained with the binaural recordings for different source positions is given in Hládek and Seeber, 2021. There, the scene rendering with the rtSOFE system (Seeber et al., 2010) is compared to the "ground truth" measurements in the real space.

## Pub

This environment is modelled after the Pub OLs Brauhaus (Rosenstraße) in the city of Oldenburg. The floor plan is shown in Fig. 3 indicating overall dimensions of about 15 x 10 m. The volume of the whole environment is about 442 m$^3$. The walls are made of plaster, the floor of oiled wood and the ceiling of raw wood supported by rough irregular wooden beams. The pub is equipped with wooden tables and chairs, and a bar desk in one corner. The resulting reverberation time is 0.7 s.

The Pub resembles an environment in which people are having social conversations and in which they can typically experience challenges understanding one another because of the babble noise and music that is playing in the background.

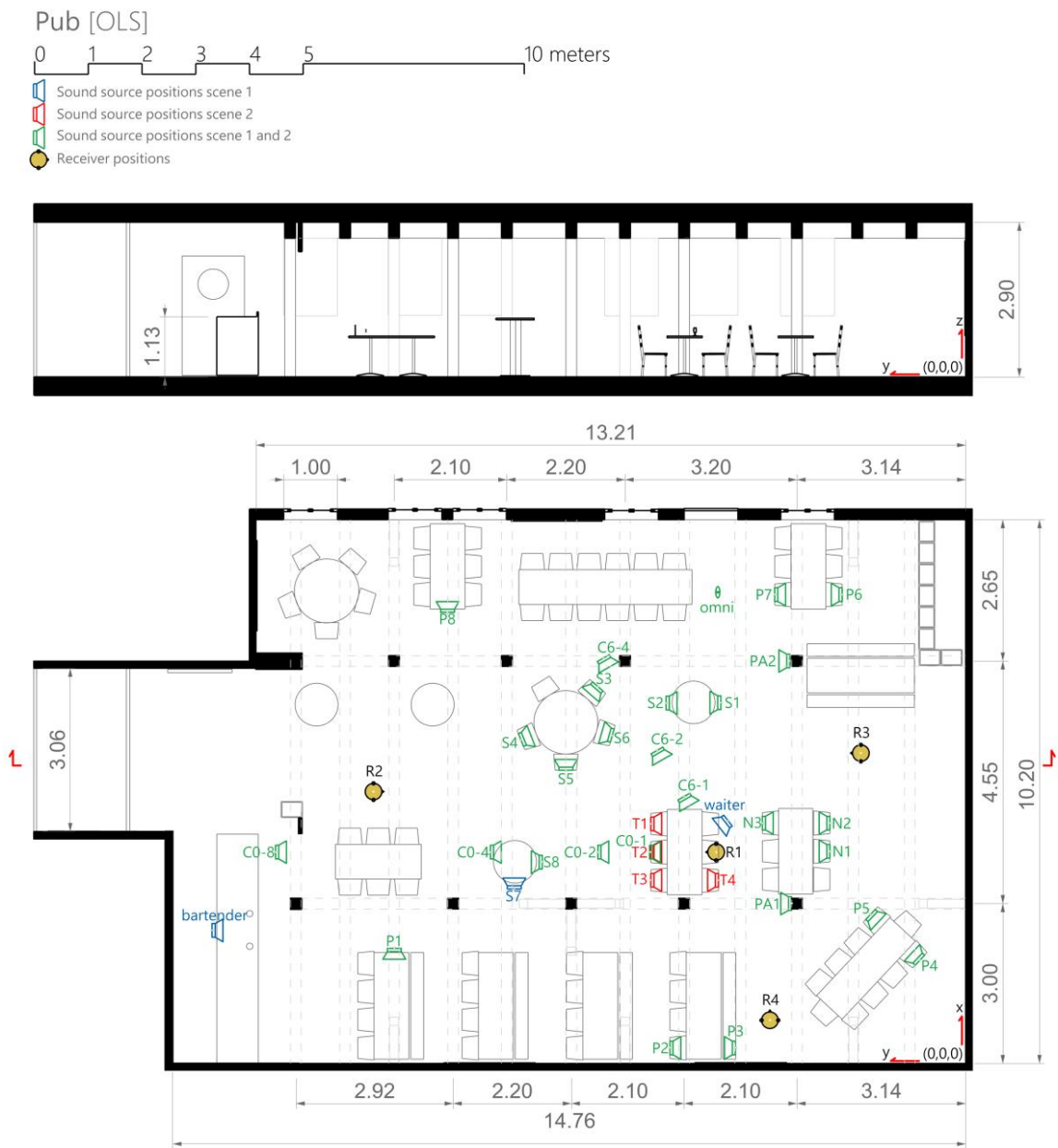

*Figure 3. Cross section (top) and floorplan (bottom) of the pub environment in the same style as in Fig. 2. Positions and orientations of receiver and sources are indicated by head, microphone, and loudspeaker symbols including their orientation in the horizontal plane. Source positions indicated in red are attributed to scene 1, blue to scene 2, green to both scenes.*

Impulse responses from many different source positions at neighboring tables were recorded, so that a babble noise background can be generated Moreover, impulse responses from a playback system for music in the pub were recorded, so that music can be added to the acoustic model. For both scenes, the receiver position R1 in Fig. 3 serves as a listener sitting at the table. Three additional receiver positions were mainly intended for the room characterization and can be used as alternative listener positions. No background sounds or near-field sounds were recorded due to privacy issues. The impulse responses recorded from the other N, S, and P positions can be used to generate background babble noise, where the speech material presented from the N positions is probably understandable because of the close proximity to the listener position. The impulse responses recorded from the PA1 and PA2 positions can be used to add background music to the acoustic environment or other nearfield sounds representative for a pub environment.

### Scene 1: Communication at a table

The first scene represents a person sitting at a table in the pub (R1) and source positions T1-4 can be used to represent a conversation with other people at the same table.

### Scene 2: Communication with the waiter/waitress

The second scene resembles the situation that a waiter/waitress comes to the table to take the order. In addition, two alternative receiver positions are provided, resembling a person standing at overall three different positions in the pub.

## Living Room

The third environment is a living room with a connected room (via a regular door) that is part of the lab infrastructure of the University of Oldenburg (main building at the Wechloy campus). The floor plan is provided in Fig. 4. The dimensions are 4.97 m x 3.78 m x 2.71 m (width x length x height), the volume of the living room is 50.91 m$^3$, while the coupled room with the dimensions 4.97 m x 2.00 m x 2.71 m has a volume of 26.94 m$^3$, resulting in an overall volume of 77.85 m$^3$. The walls are comprised of different materials: drywall material covered with wood chip wall paper can be found for three walls and the ceiling. One wall consists of bricks. The floor covering is laminate, which is partially covered by a 6 m² carpet. Window fronts are located on one side of the room, for the living room as well as for the kitchen. In the living room, a seating arrangement consisting of a textile couch and two textile armchairs can be found, arranged around a glass coffee table. On one side of the room is a cabinet, filled with glasses and decoration. Opposite from the couch there is a TV bench with a TV. To the right of the TV, there is a bookcase filled with books. In the coupled room, a table and two chairs are placed next to the wall with the window front. The reverberation time for the living room environment with the door opened between the living room and the coupled room ranges from $T_{30}$ = 0.55 s at 250 Hz to 0.46 s at 8 kHz.

The Living Room resembles an environment that people encounter frequently in their private homes and in which speech intelligibility may not necessarily be difficult – unless the television is on or the source speaker turns away from the listener. Challenging are communication situations to the connected room.

Background sounds were not recorded for this environment. Any audio from a TV show can be used for the TV source, while for the kitchen sounds of a dishwasher or a fridge are most suitable.

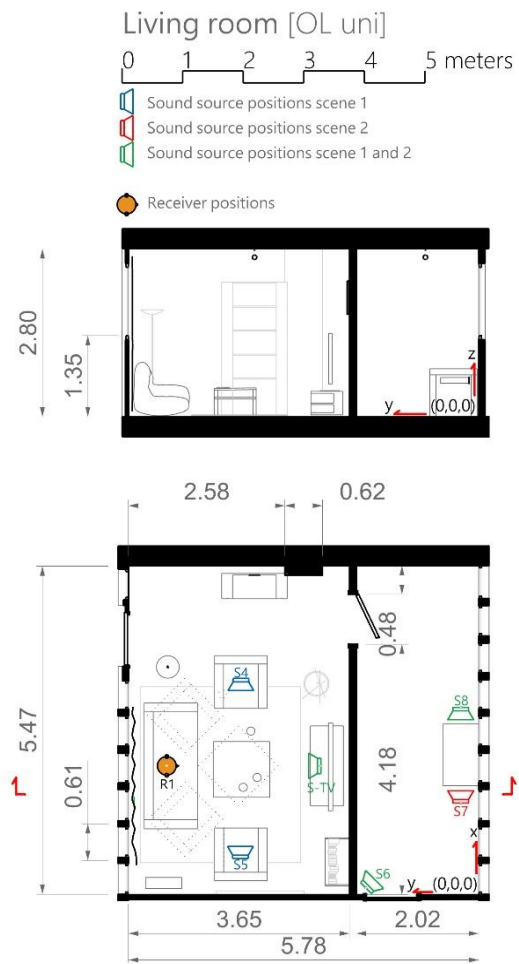

*Figure 4. Cross section (top) and floorplan (bottom) of the living room environment in the same style as in Fig. 2. Positions and orientations of receiver and sources are indicated by head, microphone, and loudspeaker symbols including their orientation in the horizontal plane. Source positions indicated in red are attributed to scene 1, blue to scene 2, green to both scenes.*

### Scene 1: Television set and communication

The listener is seated on the chair/ sofas in a conversation-like situation. The listener is listening to a second speaker, while at the same time various other sound sources are available (e.g. a television set).

### Scene 2: Communication across rooms

Here the source distance is increased and as a further aspect the listener is addressed from the neighbouring room with an obstructed direct sound path. Some recent studies suggest that such an acoustically-coupled room setting can provide extra challenges for speech intelligibility (Haeussler and van de Par, 2019; Schulte at al., 2013).

## 3. Room acoustical comparison of the environments

The provided scenes in the three environments represent different communication conditions typical for the respective environments. They follow the same principle of varying angular positions of targets and maskers in the horizontal plane (scene 1) or varying the distance to the target (scene 2), each embedded in the acoustical conditions of the environment and the respective background noise. For a comparative overview of the acoustic conditions in the three environments, Tab. 1 lists the RT 60, early decay time (EDT) and the broadband DRR for a specific source receiver condition of the two scenes side by side. In addition, for a general comparison of the different acoustics in the environments, the average values are provided and at a fixed source-receiver distance of about 1 m. As can be expected, the reverberation time is largest for the Underground station (with an average of 1.7 s) and smallest for the Living room (average 0.56 s). Because of the connected room, the reverberation time in the living room is not that much different to that of the Pub (average 0.66 s). At the fixed distance of 1 m, the DRR drops from 5.6 dB in the underground to 2.8 dB and 2.3 dB in the Pub and Living room, respectively.

Table 1. Various acoustic parameters for a comparable set of source-receiver combinations defined and measured within the three environments. In addition, room averages are calculated from a number of omnidirectional microphones and omnidirectional sources (except for the Pub, where all loudspeakers were used). Shown are the distance between source and receiver (occluded path length for the source in the adjacent kitchen of the Living room), the T30 time, the Early Decay Time, and the Direct-to-Reverberant Ratio, in columns 3 to 6, respectively.

| | Descriptor | Dist. S-R [m] | T30 [s] | EDT [s] | DRR [dB] |
|---|---|---|---|---|---|
| Underground | Room Average | | 1.68 | 0.74 | |
| | R1-S1 (Sc. 1), talker on "circle" | 1.60 | 1.11 | 0.31 | 2.7 |
| | R1-S13 (Sc. 2), talker very close | 1.01 | 1.23 | 0.01 | 5.6 |
| | R1-S15 (Sc. 2), talker somewhat distant | 4.02 | 1.25 | 0.35 | - 3.2 |
| Pub | Room Average | | 0.66 | 0.68 | |
| | R1-T2 (Sc. 1), talker on the same table | 0.97 | 0.67 | 0.17 | 2.8 |
| | R1-wtr (Sc. 2), „waiter" talking to listener | 0.90 | 0.92 | 0.46 | - 0.4 |
| | R1-C0-4, talker at medium distance | 4.00 | 0.65 | 0.56 | - 2.9 |
| Living Room | Room Average (door open) | | 0.56 | 0.46 | |
| | R1 – SG1, loudspeaker in 1 m distance | 1.01 | 0.49 | 0.13 | 2.3 |
| | R1 – S-TV (Sc. 1), television running | 2.51 | 0.49 | 0.23 | - 5.6 |
| | R1 – S7 (Sc. 2), talker in adjacent kitchen, occluded sound path | 5.69 | 0.60 | 0.25 | |

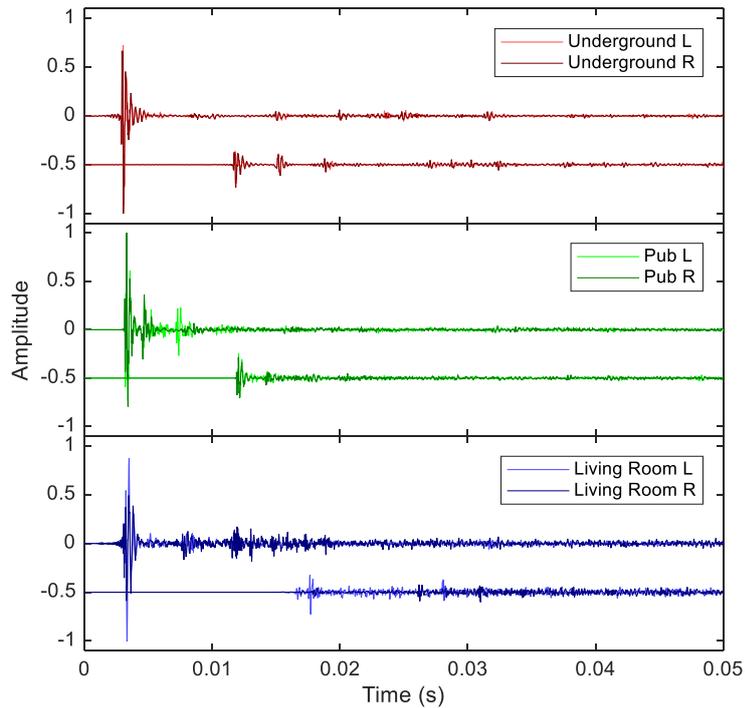

*Figure 5. Impulse responses recorded in the three environments at 1 m distance (upper traces) and at 4 m (5.6 m for the occluded path in the living room) distance (lower traces, offset by -0.5) as shown in Tab. 1. All impulse responses were normalized to the maximum of the 1-m condition. The distance-related amplitude reduction can be seen in the lower traces. Note that for better readability the lower trace in the living room lab (lower panel) was scaled up by a factor of 4.*

Figure 5 shows impulse responses recorded in the three environments at 1 m distance (top traces) and at 4 m distance (5.6 m for the occluded path in the living room) as the lower traces with an offset off -0.5, for the respective configurations provided in Tab. 1. For the Underground (upper panel) distinct reflections with relatively large temporal separation are obvious. For the larger distance (4 m), the direct sound and the first reflection (likely from the platform floor) are much closer spaced (about 3 ms) and more similar in amplitude, as can be expected. For the Pub (middle panel) a prominent early reflection from the table between source and receiver is visible for the short distance of 1 m. In the Living room (lower panel), multiple scattered and early reflections are visible reflecting the overall smaller volume with furniture. For the source in the neighboring kitchen room (lower trace), the diffracted direct sound is weaker than the first reflection which is directly followed by dense reverberation from the coupled room.

For further analysis, Fig. 6 shows the energy decay curves (EDCs) for the same conditions as depicted in Fig. 5 for 1 m (thick traces) and the larger distance (thin traces). Where the Underground (red) shows a dual-slope decay, likely related to local reverberation on the platform and a slower decay of the tube and coupled (escalator) volumes, the Pub (green) shows dominantly a single-sloped decay. For the Living room (blue) a dual-slope decay is observed for the short distance (thick trace) related to the coupled room, whereas the decay process of the coupled room dominates the condition with occluded direct sound (thin line).

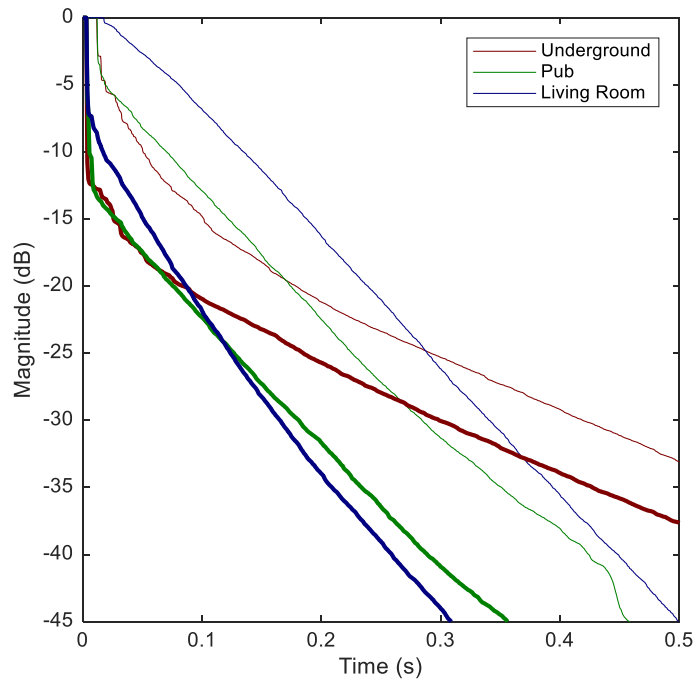

*Figure 6. EDCs for the three environments and distances (1 m, thick; 4 or 5.6 m, thin) as in Fig. 5 and Table 1.*

Taken together, the analysis demonstrates the large variety of acoustic conditions covered by the three environments presented in this contribution. Moreover, the current comparative analysis only covers a small excerpt of all conditions available for each of the environments in the accompanying dataset.

## 4. Environment description and data files

In order to create a platform for the exchange of audio-visual environments for hearing research, a recommended standard set of data is defined here to be provided for each environment (c.f. Leckschat et al., 2020). The environment and the data are described in a human readable *environment description document* (EDD), which extends the short descriptions provided above. The EDD contains all guidelines necessary to re-create the audio-visual scenes and the information provided in all files. The first part of the EDD contains all general information about the nature of the environment, the intended purpose of the scene, as well as specific data such as a floorplan, volume, and T60 times. The EDD also includes details about the recordings of background sounds and measurements conducted to obtain, e.g., impulse responses and directional characteristics of sources. The second part of the EDD describes the directory structure and file names of all provided *technical data files* (TDFs) for visual and acoustical rendering and the verification of rendering quality. Data files from specific measurements, like impulse responses, are included that allow to optimize and verify the acoustic simulation. Additionally, depending on the environment, original noise sources were recorded that can be rendered at different positions within the acoustically rendered environment, or they were recorded for direct spatial reproduction at the respective locations of receiver in multi-channel spatial or binaural audio format. The three environments with their six scenes as represented by their EDDs and TDFs have been published on Zenodo (https://zenodo.org/communities/audiovisual_scenes/).

Although for each environment two specific scenes are provided that entail sets of defined source-receiver combinations, within each audio-visual environment other combinations of the available source-receiver positions or additional positions created with virtual acoustics can be used to address specific research questions. The defined scene positions nevertheless serve as "anchor points" for reproducible research since the simulations can be compared against recorded binaural room impulse responses in the corresponding real environment and against further baseline measures taken in our labs, e.g., of speech intelligibility.

## 4.2. Environment Description Document

The Environment Description Document provides all information about the environment as structured and easily assessable human readable information. The EDD contains two sections with the following information:

ENVIRONMENT DESCRIPTION

1      OVERVIEW

      1.1      Location

      1.2      Scenes

2      FLOORPLAN

3      VOLUME

4      SOURCE-RECEIVER POSITIONS AND ORIENTATIONS

      4.1      Scene 1 – equidistant circular sources with semi-distal background noise sources

      4.2      Scene 2 – radially spaced sources

5      SCENE SOUNDS (optional)

      5.1      Single channel recordings

      5.2      Multi-channel recordings

6      ACOUSTICAL SPACE DESCRIPTION

      6.1      Reverberation time

      6.2      Direct-to-reverberant energy ratio

7      MEASUREMENT DESCRIPTION

      7.1      Measurement conditions

      7.2      Sound sources

      7.3      Measurement microphone type

Detailed information about the requirements of each of the information items can be found in the supplementary information provided on (https://zenodo.org/communities/audiovisual_scenes/).

### 4.3.   Technical Data Files

The audio-visual environments are defined in at least two separate models, one detailed model for the visual rendering, and (at least) one coarser model that allows real-time acoustic simulation and rendering. To ensure maximum compatibility, the widely used Wavefront object (.obj) format was chosen for 3D geometry which also has the advantage to be human readable and is thus easily editable with a text editor. For better usability, Blender (.blend) files as well as Unreal engine projects are optionally provided. For the visual model, textures are provided in separate files referenced in the .obj files or in accompanying material .mtl files. For simplicity, only simple (colormap or albedo) textures are provided. For the acoustic model, absorption (and if available scattering) properties are provided in a separate .txt file linked to the material names referenced for the surfaces in the acoustic .obj model.

For the acoustic model, the level of detail should be sufficiently accurate to allow a faithful simulation of the room acoustics. On the other hand, the level of detail should be sufficiently low to allow for a reasonable computational effort in simulating the room acoustics in a real-time system. The minimum requirement for a faithful rendering is not precisely known. Based on our experience with current real-time room acoustical rendering software, the following is proposed:

- Use a total of about 25 surfaces to define the boundaries of the acoustic environment simulated.

- Simulate additional surfaces close to sources and receivers when they create a respective spatial opening angle to the source or receiver that is more than 36°x36° (this entails about 1% reflective surface of a full 4π spatial angle).
- Use expert insight about the particular simulated environment to finalize the level of detail required for the simulation.

For sources, directionality data can be provided in SOFA-format (AES69-2015). This is of specific relevance for the acoustical room simulation to be able to compare measured impulse responses to simulated impulse responses. In addition to source directivity, receiver directivity can be included in SOFA-format. Specifically, for the measurements made with an artificial head, the availability of a full set of head-related transfer functions (HRTFs) will allow comparing the measured binaural room impulse responses (BRIRs) with the simulated BRIRs using the acoustical simulation and rendering method the user chooses.

Linked to the specified source-receiver positions, a set of measurements is provided that were made in the real location to allow for optimization and verification of the acoustic rendering. The measurements support three purposes:

The first is that BRIRs are measured for at least one specific source-receiver combination. Once the specific scene is simulated and rendered, it can be inspected whether the auralization matches the measured BRIRs. In this case, optimally, headphone rendering should be used, while that auralization is performed with an HRTF-set that is measured with the dummy head used for the BRIR measurements.

The second is that room impulse responses allow determining frequency dependent T60 times. Based on T60 times, the acoustic simulation of an environment can be optimized for example by adjusting absorption coefficients of the environment's surfaces. For measuring T60 times we recommend generally following ISO – 3382-2:2008, possibly with a reduction in the number of measurements such as using at least one source and two receiver positions.

The third is that, ideally, recordings of typical scene background sounds are provided. These scene sounds would specifically entail background sounds that would be regarded as interfering sources in a communication scenario. These scene background sounds can be recorded in two manners. One manner is that recordings are made near a background source (or a number of background sources). In this case the recordings can be rendered as, effectively, anechoic recordings that are placed in the scene at their respective places, and room acoustics is added as part of the simulation. The second manner would be to capture the spatial sound field with a multi-channel microphone for reproduction via, e.g., higher-order Ambisonics, which allows the rendering of the captured spatial sound field either via headphones via a loudspeaker set up.

Besides the above recordings, also calibration files are provided. These are recorded signals of a calibrator placed on the measurement microphones in order to relate the recorded signals to a specified sound pressure level.

### 4.4. Contributions to the Environment Data Set

The Zenodo channel on which the AV environments have been made publicly available is open for contributions from the community via the Zenodo channel https://zenodo.org/communities/audiovisual_scenes/. Potentially, some contributions will be made that entail environments that are of particular research interest, but that do not exist in real life. In this case the corresponding measurement data will not be obtainable and can be discarded as part of the environment description document and as part of the environment data files. A template of the Environment Description Document with further submission information is provided in the Zenodo channel.

# 5. Discussion

A framework was presented to define audio-visual environments that can be used for hearing research in complex acoustical environments. The framework contains visual and acoustical models that can be rendered with room acoustical simulation methods and visual rendering engines. In addition, each environment is supplemented with a range of measurements that allow to optimize and verify the acoustic rendering. Within each of the presented three environments two 'scenes' are defined which represent specific source-receiver combinations that are typical for such an environment. Furthermore, sound recordings of background sources are included that are typical for such an environment. The presented framework, which can be retrieved via an on-line repository (https://zenodo.org/communities/audiovisual_scenes/), is open for future contributions from the general scientific community.

The implementation of the current environments in acoustic and visual virtual reality rendering engines requires special attention to achieve reproducible results in auditory-visual research studies. For this reason, the Environment Description Document available for each environment provides detailed guidelines regarding the process of rendering the environment with the acoustic and visual models, namely: origin reference position, Z orientation, normal face orientation, material IDs, visual textures, receiver location and direction, source location and direction. The document further details the content of the Technical Data Files which contain the information for rendering, e.g. the acoustical models and impulse responses measured at scene positions.

Related efforts regarding virtual acoustic environment emerge in the literature. Brinkman et al. (2019) conducted a round robin to evaluate the state-of-the art in room acoustical simulation and auralization. In this study, the focus was on evaluation of various existing simulation methods in terms of technical accuracy, and in terms of perceived authenticity. Llorca-Bofí et al. (2018) investigated the use of 3D photogrammetry to support acoustic measurements and to derive geometries for simulation. Their work relies on the photographic data – from concert halls, auditoriums and theatres-, to extract geometric information by triangulation algorithms, as well as acoustic material definition of surfaces objects via deep learning methods.

The present work will support future research into different rendering methods and their suitability to assess hearing abilities in more real-life complex environments. Further extension will also be needed when users can interactively move within the provided environments. The acoustic model can be extended to have a variable level of detail to be able to incorporate the variable effect close-by objects have on the perceivable sound field. For example, the underground scene's acoustic models are provided in three versions differing in detail. Related to this, Llorca-Bofí and Vorländer (2021, a, b, c), published a multi-detailed 3D architectural framework for sound perception research in Virtual Reality.

### Future relevance for hearing research
The presented framework is envisioned to strengthen research on speech intelligibility and more general hearing research. The availability of ground truth data for each complex acoustic environment will allow verifying speech intelligibility in the acoustic simulations, permitting to make stronger assertions based on the findings of experiments performed in such virtual environments. In general, the virtual environments will allow to obtain subjective data in contexts more similar to real-life. Currently surveys are exploring situations and environments in which persons with reduced speech recognition ability experience most challenges (e.g., hearing impaired listeners, Gablenz et al., 2021). With the proposed framework, these environments and situations can be used in lab-based experiments and will be available across laboratories to allow to deepen understanding on speech intelligibility in such complex environments within the scientific community.

The framework provided here will allow active participation of subjects in the environment. More specifically, the role of head movements in response to an active conversation can be investigated. Factors revolving about auditory attention, supported by visual information can be taken into account within a complex environment that is relevant in daily life. The interaction of head-movements with hearing aid processing can be studied within an audio-visual environment that should elicit much of the typical head-movement behaviour that would also be observed in daily life (Hendrikse et al., 2020).

Having a virtual acoustic rendering of complex acoustic environments, will allow to specifically manipulate auditory cues to get a better understanding about their relevance. Factors such as conservation of spatial ITD, ILD, and IACC cues in binaural hearing aids can be investigated in relevant daily-life settings. In addition, it is possible within such a virtual environment to create the 'perfect' hearing aid, that amplifies a single source, even in an interactive setting.

Complex auditory-visual environments that simulate every day settings will allow to better probe cognitive factors involved in processing speech information by (hearing impaired) listeners in daily life. It stands to reason that the complexity of every day acoustic environments will be of relevance for the way cognitive resources are used by hearing impaired listeners.

Finally, also for more basic hearing related questions, such as the precedence effect, perception of moving sound sources, and distance perception, complex acoustic environments provide a means to gain better understanding about the perceptual mechanisms underlying these perceptual phenomena specifically within daily-life complex settings.

# Acknowledgement

This work was supported by the DFG Collaborative Research Centre SFB 1330 Hearing Acoustics (HAPPAA), projects B1, C2, C4, and C5.